# TIMGen: Temporal Interest-Driven Multimodal Personalized Content Generation

Tian Miao

School of Artificial Intelligence and Automation, Huazhong University of Science and Technology, Wuhan, Hubei, China

tian_miao0001@hust.edu.cn

**Abstract**: With the dynamic evolution of user interests and the increasing multimodal demands in internet applications, personalized content generation strategies based on static interest preferences struggle to meet practical application requirements. The proposed TIMGen (Temporal Interest-driven Multimodal Generation) model addresses this challenge by modeling the long-term temporal evolution of users' interests and capturing dynamic interest representations with strong temporal dependencies. This model also supports the fusion of multimodal features, such as text, images, video, and audio, and delivers customized content based on multimodal preferences. TIMGen jointly learns temporal dependencies and modal preferences to obtain a unified interest representation, which it then generates to meet users' personalized content needs. TIMGen overcomes the shortcomings of personalized content recommendation methods based on static preferences, enabling flexible and dynamic modeling of users' multimodal interests, better understanding and capturing their interests and preferences. It can be extended to a variety of practical application scenarios, including e-commerce, advertising, online education, and precision medicine, providing insights for future research.

# 1 Introduction

## 1.1 Background

With the continuous development of the digital age, user interest and behavior data have gradually been regarded as important resources in internet applications. How to provide personalized content generation to users in a changing environment has become a common concern of academia and industry [1] . Most of the existing work uses traditional static modeling methods to interactively model users' historical interactions to capture user preferences and thus generate personalized content. However, these works are limited to modeling static interests or long-term trends, and ignore the consideration of the transfer of interests over time, context, and external events [2] . In a complex Internet environment, it is necessary not only to characterize the evolution of user preferences over time series, but also to consider the use of appropriate multimodal weight control adjustments to dynamically adapt to immediate interests and contexts [3] . If a balance cannot be established between these two aspects, the user experience will be degraded. For example, the e-commerce industry combines users' immediate purchase intentions and historical behaviors to quickly generate personalized product recommendation results to improve conversion rates and retention rates; in digital advertising, accurate real-time content push and reasonable return on investment are required, which is closely related to efficient dynamic adaptation of user interests [4] . Therefore, content generation technology based on static interests often finds it difficult to meet the needs of diversity and high-frequency changes [5] .

With the deepening development of the digital age, user interest and behavior data are increasingly being developed and utilized as crucial resources in internet applications. Generating content for users in this ever-changing environment has attracted widespread attention from both academics and industry. Existing methods often build user models based on static descriptions of historical interactions. These methods struggle to capture dynamic changes, especially in the rapidly changing and complex landscape of internet applications, where user interests are likely to rapidly shift over time, context, and external events. Therefore, it is crucial not only to consider how to leverage the temporal nature of user interaction histories to describe the evolution of user interests but also to flexibly adapt to changes in immediate interests and the external environment. For example, in e-commerce scenarios, it is often necessary to rapidly generate a certain level of product information based on users' real-time interests and historical behaviors to drive conversion and retention. Similarly, in digital advertising, content generation mechanisms must be able to adapt their targeting strategies to specific audiences, ensuring maximum return on investment and effective ad targeting.

Therefore, content generation based on static interests struggles to cope with diverse and frequently changing demands. This paper proposes the TIMGen (Temporal Interest-driven Multimodal Generation) framework to address the rapid evolution of user interests and their incompatibility with multimodality. Compared to approaches that primarily consider static modeling, this framework provides a more flexible approach to long-term personalized multimodal generation and lays the foundation for further characterizing momentary interest fluctuations.

## 1.2 Problem Description

interest modeling in personalized content generation mainly predicts the content that users may be interested in in the future based on their entire historical behavior data over a certain period of time. This problem can be described as follows:

Assume that the user $u$'s historical interaction sequence is:

$$H_u = \{x_1, x_2, \ldots, x_t\}$$

where $x_t$ represents the user's interaction over time $t$, and the objects they interact with have multimodal characteristics:

$$x_t = \{x_t^{text}, x_t^{img}, x_t^{video}, x_t^{audio}, \cdots\}$$

The user interest representation constructed by the traditional personalized generation method is:

$$z_u = f(H_u)$$

where $f(\cdot)$ is a static feature extraction function that generates a single preference vector ($z_u$).

However, the above traditional models are difficult to effectively deal with the rapid evolution of user preferences within a future window period, which is mainly manifested in three aspects: (1) Insufficient temporal dynamics, which can only describe the overall preferences of the whole world, but cannot describe the immediate transient interests; (2) Insufficient multimodal adaptation, which ignores the importance differences of different modalities in different situations and is difficult to fully reflect interests; (3) Insufficient dynamics leads to insufficient personalization, which fails to describe the dynamic changes of individual interests. Therefore, the generated result categories are limited and homogenization problems are very likely to occur. The TIMGen method fully considers the temporal characteristics in the interaction sequence, introduces the temporal function $g(\cdot)$, models the user's interest status at each time $t$, and integrates multimodal features into the sequence modeling process. That is, the following problem is constructed:

Assume that the user $u$'s historical interaction sequence is:

$$H_u = \{x_1, x_2, \ldots, x_t\}$$

where $x_t$ represents the user's interaction over time $t$, and the objects they interact with have temporal features, multimodal features, and other features:

$$x_t = \{t_t, x_t^{text}, x_t^{img}, x_t^{video}, x_t^{audio}, \cdots\}$$

Dynamic interest representation is obtained through temporal feature extraction function $g(\cdot)$:

$$Z_u = \{z_1, z_2, \ldots, z_t\}, z_t = g(x_1, x_2, \ldots, x_t)$$

The user's interest state at a certain time $t$ is $z_t$, not only dependent on the multimodal feature sequence, but also constrained by the real-time interactive operation. At time $t$, during the personalized generation process, the system generates personalized content $\hat{y}$ for the candidate items $i_t$. The generator is $G(\cdot)$ conditioned on the recent interests $z_t$ and target item features $x^*$:

$$\hat{y} = G(z_t, x^*)$$

The research objective is further stated as: under the condition of a given historical sequence $H_u$, construct dynamic interests $z_t$ and effectively integrate them into the generator $G(\cdot)$, so that the generated content $\hat{y}$ takes into account both item relevance and user personalization.

## 2 Related Work

How to model user interests is a key issue in personalized content generation. Early methods modeled user interests as static vectors, such as using matrix decomposition and collaborative filtering to generate fixed interest preference representations. Since user interests are dynamic and temporal, static interest modeling methods are difficult to capture the trend of interest changes over time. To solve this problem, some work explored how to learn users' long-term and short-term interests. K.Sun et al. [6] used two different interests, long and short, to recommend POIs, and verified that focusing on users' recent interests can improve the accuracy of POI recommendations . Y.Wu et al. [7] further proposed using LSTM and attention mechanisms to simultaneously learn users' short-term behavior and long-term preferences, and achieved better results than the original model on multiple data sets . P.Zhao et al. [8] proposed multi-interest and multi-scenario joint modeling in the M5 framework, further improving the recommendation performance in OTT scenarios. This type of method has improved timeliness and personalization, but it is often targeted at a certain type of information source and difficult to extend to multimodal scenarios.

In personalized recommendation and content generation tasks, in order to ensure that the content has a high degree of acceptance, multimodal signals such as text, images, and videos should have different focuses in different scenarios. Therefore, more and more studies have begun to explore the significance of multimodal features for personalized modeling. The review article by Q. Liu et al. [9] found that various types of information of varying degrees, such as vision, language, and interaction, can better capture and characterize user personality preferences . A. Kristensen et al. [10] proposed a diffusion model quantification method based on large-scale generative AI. Through an efficient quantification strategy, the performance of multimodal conditional generation was improved, which can better support personalized content generation tasks, especially in applications under large-scale data sets. Z. Wang et al. [11] applied the multimodal large language model (MLLM) to long sequence recommendations, further improving the model's ability to model complex interests . J. Wang et al. [12] used knowledge graphs for entity and semantic mining, and integrated multimodal information enhancement capabilities . The MISSRec framework proposed by J. Wang et al. [13] transfers modal knowledge from a multimodal pre-trained model and introduces external semantics, achieving interest-aware long-term sequence modeling based on multimodal signals. This solves the cold start problem while enhancing the model's ability to generalize multimodal recommendations. These studies have shown that multimodal signals can help improve the diversity of recommendations and generation, but most work only weights different modalities, making it difficult to adaptively adjust the importance of different modalities based on the temporal changes in user interests.

In terms of using generative AI for personalized content generation, with the rise of large language models (LLMs) and multimodal large models (VLMs/MLLMs), research has shifted from recommendation to generation. Shen et al. [14] proposed the PMG framework, using LLM for personalized multimodal generation, but the interest modeling in it still remains in the traditional static aspect . Liu et al. [15] proposed "Generate, Not Recommend", pointing out that the powerful ability of LLM can directly generate multimodal content that is currently relevant to the user, and it is no longer a recommendation problem . At the same time, they pointed out the decisive role of long-term interest modeling in the generation quality. Wei et al. [16] used a large-scale visual-language model to achieve cross-modal generation of personalized recommendations, demonstrating the potential of unified personalized generation. The review article by Xu et al. [17] summarized the personalized generation trend in the era of large models and pointed out that temporal modeling and multimodal interaction are still the current bottlenecks. Although the existing generation methods all use the super-strong cross-modal generation capabilities of large models to generate personalized results, they are still unable to capture the user's immediate interests.

shows the following development trends in interest modeling, multimodal fusion, and content generation:

（1） The evolution of interest modeling: from static user vector modeling to joint interest modeling that can take into account the connection between users' short-term behavior and long-term preferences and consider them together.
（2） Expansion of modal range: gradually expanding from single modality generation to multi-modal fusion generation.
（3） A shift in research paradigm: from recommending other items within a given range to directly providing users with pictures, texts, audio and video.

However, existing methods still have shortcomings: (1) It is difficult for the model to capture and process the dynamic changes of users' interests; (2) The information fusion between different modalities is often carried out through static or weak dynamic weighting, which lacks the ability to adjust according to the actual scenario; (3) There is a lack of rapid response to users' short-term interests. Therefore, a personalized multimodal generation method with learnable modal adaptive weights that can capture changes and establish dynamic interests over a long time series is needed to meet the personalized needs in complex situations.

## 3 Review of MISSRec

### 3.1 Summary of MISSRec

J.Wang et al. [13] proposed MISSRec ( Multi-modal The MISSRec method proposes a multimodal, interest-aware sequential recommendation framework. The core of the MISSRec approach is to incorporate information beyond IDs into user interactions, overcoming the cold-start and transfer weaknesses of traditional collaborative recommendation systems based on IDs. MISSRec designs a Transformer-based context encoder to capture the relationships between different types of information and proposes an interest discovery module to encode user multimodal information into a set of interest tokens, removing redundancy and enhancing sparse yet important interest signals. Furthermore, an interest-aware decoder is used to model the relationship between items, modalities, and interests, enhancing the expressive power of sequential representation learning. Furthermore, MISSRec introduces a lightweight dynamic fusion module to represent candidate item information. Using model pre-training techniques based on contrastive learning, simple fine-tuning techniques are used to achieve good transfer performance across domains, enabling MISSRec to improve model generalization and application in new scenarios.

### 3.2 Strong Points

MISSRec has made significant contributions to the field of personalized multimodal recommendation. Its main advantages include:

- Addressing Cold Start and Sparsity: This approach addresses the severe cold start and sparsity issues faced by traditional ID-based applications, such as e-commerce and content recommendation systems. This approach leverages the interaction between multimodal information, contextual information, and interests to capture more useful information and enables MISSRec to be transferred to new tasks.
- Interest modeling innovation: The combined design of "interest discovery module + interest-aware decoder" removes redundancy while capturing fine-grained item-modality-interest relationships.
- Complete architectural design: The framework integrates multimodal feature adapters, contextual Transformer encoders, and dynamic fusion mechanisms, with a certain degree of engineering and technical depth.
- The results are fully verified by experiments: a comprehensive evaluation was conducted on multiple real-world datasets, showing improved cold start performance in different fields and scenarios, with results that outperform current mainstream recommendation methods.

### 3.3 Weak Points

Although the MISSRec method has shown good performance in multimodal personalized recommendation, it still has some shortcomings:

- Lack of dynamics: This method primarily relies on interest tokens to reflect users' preferences or interests, but fails to model and characterize users' complex temporal dynamics, making it difficult to demonstrate the long-term evolution of users' interests.
- Complexity and efficiency: The framework consists of multiple modules (context encoder, interest discovery, dynamic fusion, and contrastive learning). Although it can achieve higher-quality recommendation accuracy, it also brings high computing and training overhead, which is not conducive to large-scale real-time deployment.

- Limited experimental scenarios: Due to the scale and scope of the problem, the evaluation mainly tested the effectiveness of the model based on public datasets, but lacked online testing on large-scale industrial systems.
- Insufficient explainability: Adding interest tokens to the model improves the model's effectiveness, but the correspondence between interest tokens and users' actual behavior patterns has not been fully explained and analyzed.

### 3.4 Detailed Analysis

The following review comments are made from the perspective of paper writing and presentation:

- Novelty is not prominent enough: In the related work, there needs to be more direct comparison with existing work on sequence recommendation in other modalities. There is little comparison with existing multimodal large-scale models, which makes the research work somewhat less localizable.
- Insufficient details: Some key details (such as the interest token generation process) are briefly described, which may affect the reproducibility of the method. It is recommended to provide a complete formula derivation or pseudocode to further improve the clarity, reproducibility, and generalizability of the method.
- The framework diagram has limited readability: Fig.2 (framework diagram) is too information-intensive, and the information and connection relationships are not easy to see at a glance. You can consider using a hierarchical structure (such as input layer - representation layer - interest modeling layer - output layer) to draw it.

## 4 Review of PMG

### 4.1 Summary of PMG

Shen et al. [14] proposed the PMG (Personalized Multimodal Generation) framework, applying large language models to personalized multimodal generation. The core idea of PMG is to convert user interaction records such as clicks, conversations, and operations into natural language format (explicit text information), thereby learning the user preferences contained in the explicit text information. PMG uses explicit keyword representation and implicit embedding representation to represent user interests. It then uses the obtained preference information combined with keywords related to candidate items as conditions to guide large generative models (such as diffusion models or multimodal large language models) to output personalized output results. The PMG method designs a weighted objective function to jointly optimize the generative model to achieve a balance between relevance and personalization.

### 4.2 Strong Points

The advantages of the PMG method are mainly reflected in the following aspects:
- Application Value: This technology extends personalization from "recommendation" to "generation," transforming recommendation systems into multimodal generation systems.

It also demonstrates initial application prospects in advertising, social media, and e-commerce.
- Preference modeling innovation: The combination of "keywords + implicit embedding" compensates for the shortcomings of natural language expression and improves the sophistication of preference expression.
- Improved method: The method includes preference extraction and content embedding. Through modeling and optimization training of balanced conditions, a more efficient generative model is obtained. It is combined with P-Tuning v2 and tokens to achieve generative adaptation of large models.
- Comprehensive Experimentation: We conducted comprehensive experiments and analysis on clothing, movies, and facial expressions, achieving promising results. Our model outperformed the control model on various metrics (LPIPS, SSIM, and manual evaluation).

### 4.3 Weak Points

Although PMG shows good application prospects, there is still room for improvement in terms of dynamics, generation quality and efficiency:
- Insufficient interest dynamics: Preference modeling mainly relies on static historical interactions and lacks the characterization of long-term evolution and immediate interest changes.
- Limited generation quality: For example, the characters on the generated movie posters are not real actors, and the pictures of clothing do not match the actual purchased products.
- Limited efficiency: The framework is an LLM cascade diffusion model, and the inference time is too long (generating a single image takes about 5 seconds in the experiment), and it cannot respond in real time.

### 4.4 Detailed Analysis

PMG is a significant exploration in the field of personalized multimodal generation. It proposes transforming user behavior into signals understandable by LLM and combining keywords with embeddings to model user preferences, achieving a balance between interpretability and expressiveness. Based on the paper's content, the following two issues and suggestions emerge during the reading process:
- The comparative research is not systematic enough: Although the paper reviews some fields related to this work, including GANs and traditional machine learning methods in the recommendation field, LLM-based recommendation, etc., it does not provide a very thorough comparison of the differences with PMG, especially the unique advantages of PMG compared to other algorithms.
- The method details are not clearly stated: the formulas for some key details (such as the training process of soft preference embeddings) are not intuitive and are not written together. You need to read the appendix to get the full picture, which is difficult to understand for researchers in non-generative fields.

# 5 Proposed Method

## 5.1 Overall Design

The TIMGen method is a long-term interest modeling and personalized multimodal generation framework that can capture the dynamic evolution of user interests and adaptively adjust modal weights through a multimodal interaction mechanism, thereby taking into account both long-term preferences and short-term interests in the generated results.

In terms of dataset construction, this study integrates user behavior sequences with multimodal information, encompassing text, images, video, and audio signals. It also explicitly introduces temporal features to achieve dynamic interest modeling. Labels are designed in both "personalized scoring" and "categorization" formats, reflecting user preference strength and interest direction, respectively, to provide more accurate interest judgment and content selection.

Algorithmically, TIMGen uses long-term interest representation as its core, combining temporal interest modeling with multimodal interest preference modeling to form a comprehensive representation of user interests. This representation is then mapped into a latent space using a generative model and decoded to obtain content that matches individual interests. This objective function combines reconstruction constraints with label-based supervision.

## 5.2 Dataset Collection

### 5.2.1 Collection method

In order to ensure the reproducibility of existing data and make up for the deficiencies in multimodality and temporal dynamics, a hybrid approach of "public data set + platform data collection" is adopted.

（1） Public datasets
- MovieLens-1M: Provides millions of rating records, suitable for long-term time series modeling and verification. However, it lacks multimodal features such as images and videos and only has text and rating information. Therefore, it serves only as a baseline for long-term time series.
- Amazon Review: This data provides user reviews and product images from various fields, making it suitable for text-image fusion experiments. However, it lacks video information, and interest changes rapidly. Therefore, this paper primarily uses this data for multimodal comparative verification experiments.
- MM-Rec dataset: Contains multimodal features such as text and images, and can be used to support multimodal feature data. However, the data capacity provided in this dataset is limited, and its interaction time is short. Therefore, it is only used for supplementary experiments.

（2） Platform data collection

To address the modality gaps in publicly available data, we further acquired a large amount of user behavior and data across various modalities through APIs such as IMDb, Twitter, and YouTube, with a particular focus on supplementing video and audio content. Where relevant data was missing, we used crawler technology to capture reviews and product pages, public social media information, and some images. This data was then supplemented using feature alignment and semi-supervised methods, ultimately resulting in experimental data consisting of text, images, video, and audio.

**5.2.2 Feature Selection**

TIMGen selects features from three perspectives: user behavior features, temporal features, and multimodal features. User behavior features ensure direct and differentiated interest characterization, temporal features ensure that interest representations can evolve dynamically over time, and multimodal features expand the expressive dimensions of interest modeling.

（1） User behavior characteristics

User interactions are the most fundamental signals for interest modeling. Interactions such as clicks, browsing, purchases, likes, and comments not only reflect a user's level of interest in an item, but also the intensity of their expressed interest—both explicit and implicit. For example, clicks and browsing often reveal superficial user preferences, while purchases and comments reveal deeper levels of interest.

Furthermore, user behavior is often influenced by context, such as the device used for interaction, the platform on which the interaction occurs, and the geographic location. This contextual information helps reveal user habits and differentiated needs in different contexts. For example, the reading experience for the same user on a PC and a mobile phone may be completely different, or people in different locations may have different cultural characteristics. By using context as a characteristic of user behavior to help describe interest models, content generation can be more tailored to the user's current context.

（2） Timing characteristics

User interests evolve dynamically over time. Therefore, temporal features are introduced. Absolute time features (such as date and hour) can characterize user activity habits in different time periods. Time interval features can reflect the correlation between two user behaviors; the closer the two actions are, the more similar the user interests they reflect. Furthermore, periodic features (such as weekly patterns and seasonal preferences) can reveal the repetitiveness and rhythmicity of user interests. For example, a user may choose to consume or entertain on weekends, but be more likely to study or use other tools on weekdays. By modeling temporal features, the model can more sensitively capture the dynamic changes in interests.

（3） Multimodal features

Multimodal data such as text, images, videos, and audio can complement and integrate each other to comprehensively reflect user interests.

- Text data (such as comments and search records) can reflect users' explicit semantic needs;
- Image data （such as product images and video screenshots） can reveal users' visual preferences;
- Video data (such as movie clips and short videos) has a concept of time and can provide richer interest signals;
- Audio data (such as music or video soundtracks) can provide more information about hearing.

### 5.2.3 Tag Definition

The intensity of user interest can generally be reflected through behavioral indicators such as ratings, click-through rates, or dwell time. These rating labels can capture fine-grained differences in interest, enabling the model to learn varying degrees of preference, thereby enhancing the personalization of generated results. At the same time, user preferences in multimodal scenarios often manifest as significant categorical features, such as movie genres, image styles, or product categories. Categorical labels provide clear semantic constraints for the generation process, enhancing the relevance and controllability of the content. Therefore, TIMGen employs a dual form of rating labels and categorical labels. The former measures the strength of the user's preference for the generated content, denoted as $y_t^{score}$; the latter indicates the category or style of the generated content, denoted as $y_t^{class}$.

（1） Personalized scoring label definition

Assume that the preference intensity corresponding to the interaction between the user $u$ and the item $i_t$ at time $t$:

$$y_t^{score} \in \mathbb{R}$$

➤ In the movie recommendation scenario:

Preference can be directly portrayed through movie reviews:

$$y_t^{score} \in \{1,2,3,4,5,6,7,8,9,10\}$$

➤ In the e-commerce scenario:

It can be modeled through multi-dimensional behavioral signals, including click, add-to-cart, purchase, and comment.

$$y_t^{score} = \alpha \cdot I_{click}(i_t) + \beta \cdot I_{cart}(i_t) + \gamma \cdot I_{purchase}(i_t) + \delta \cdot I_{comment}(i_t)$$

- $I_{click}(i_t) \in \{0,1\}$: Whether the user clicks
- $I_{cart}(i_t) \in \{0,1\}$: Whether the user adds to the shopping cart

- $I_{purchase}(i_t) \in \{0,1\}$: Whether the user purchased
- $I_{comment}(i_t) \in \{0,1\}$: Whether the user comments

The weight parameters $\alpha$, $\beta$, $\gamma$, $\delta \in \mathbb{R}^+$ can be set according to the business scenario.

➢ In the video recommendation scenario:

You can use viewing time, viewing ratio, interactive behavior, dwell time, etc. as labels:

$$y_t^{score} = \lambda_1 \cdot \frac{w_t}{L_t} + \lambda_2 \cdot I_{like}(i_t) + \lambda_3 \cdot I_{comment}(i_t) + \lambda_4 \cdot I_{share}(i_t)$$

The total viewing time of the video $i_t$ is $w_t$, $\frac{w_t}{L_t}$ is the viewing completion rate.

$I_{like}(i_t) \in \{0,1\}$: Whether the user likes the post

$I_{comment}(i_t) \in \{0,1\}$: Whether the user comments

$I_{share}(i_t) \in \{0,1\}$: Whether the user shares

The weight parameters $\lambda_1$, $\lambda_2$, $\lambda_3$, $\lambda_4 \in \mathbb{R}^+$ can be adjusted according to the application objectives.

(2) Category tag definition

Suppose the category label of the generated content is

$$y_t^{class} \in \{c_1, c_2, \ldots, c_k\}, k < K$$

Where $K$ represents the total number of categories and $c_k$ represents the $k$-th category. The definition of $y_t^{class}$ in a typical scenario is as follows:

In the movie recommendation scenario, $y_t^{class}$ take a set of movie genres (such as Action, Comedy, Romance, etc.);

In image generation, $y_t^{class}$ take a set of image styles (such as Realistic, Cartoon, Illustration);

In e-commerce recommendations, $y_t^{class}$ take a set of product categories (such as Fashion, Electronics, Home).

$y_t^{class}$ use one-hot vector or trainable embedding representation:

$$e_{class}(y_t^{class}) \in \mathbb{R}^{d_c}$$

## 5.3 Model and Algorithm

### 5.3.1 Algorithm Principle

The core concept of TIMGen is that user interests evolve over time and are expressed in various forms (text, images, videos, etc.). As shown in Fig 1, the overall TIMGen process consists of five layers: input encoding layer, temporal interest modeling layer, multimodal fusion layer, personalized generation layer, and output layer. At the input encoding layer, the model extracts feature from various behavioral data related to user interests, achieves spatial alignment between these types of behavioral data, and constructs a unified sequence representation $H_u$. In the temporal interest modeling layer, the Transformer architecture is used to model interest information in the time dimension, capture different types of user interests and their long-term relationships in the evolution process, and obtain temporal interest representation $z_t$; the multimodal interest fusion layer introduces a learnable self-attention mechanism to dynamically allocate the contributions of different modal features at different times under a given context, obtain modal-level interest representation $z_t^{multi}$, and combine it with the temporal interest representation to form the final interest vector $z_t^{final}$; in the personalized generation layer, the interest vector is encoded into the latent interest space through the variational autoencoder (VAE), and the latent variables are sampled by reparameterization $\ell$ to generate content consistent with user preferences; in the result optimization stage, the reconstruction and regularization constraints of $\mathcal{L}_{VAE}$ are comprehensively considered, the label supervision signals of $\mathcal{L}_{score}$ and $\mathcal{L}_{class}$ are used to construct a joint loss function, so that the model generation results can meet the user's interests and match the semantic category.

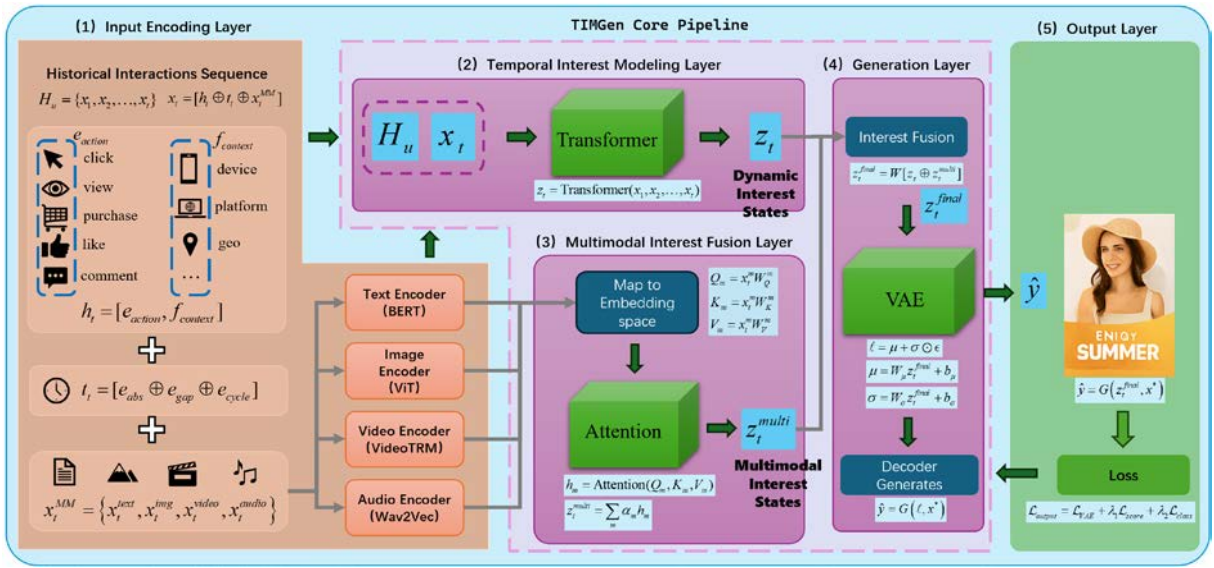

Fig 1 Overview of TIMGen Framework

### 5.3.2 Input encoding layer

To effectively characterize the long-term evolution of user interests and support personalized multimodal generation, TIMGen's input data is the complete historical interaction information $H_u$ of a single user $u$:

$$H_u = \{x_1, x_2, \ldots, x_t\}$$

where $x_t$ represents an interaction. Each interaction is characterized by three types of features: user behavior features $h_t$, temporal features $t_t$, and multimodal features $x_t^{MM}$:

$$x_t = [h_t \oplus t_t \oplus x_t^{MM}]$$

（1） User behavior characteristics $h_t$

User behavior characteristics mainly describe the user's actions in a certain context, such as clicks, browsing, purchases, likes, and comments. Assume that the user behavior characteristics $h_t$ can be expressed as:

$$h_t = [e_{action}, f_{context}]$$

The embedding of interaction type is represented by $e_{action}$ and the embedding of context information (such as interaction device, interaction platform, geographic location, and interaction time) is represented by $f_{context}$.

Interaction features $e_{action}$ include five categories:

- Click
- View
- Purchase
- Like
- Comment

After one-hot encoding, it is mapped to a $d_a$ dimensional vector space through a trainable embedding matrix $E_{action} \in \mathbb{R}^{5 \times d_a}$:

$$e_{action} = E_{action}[type(i_t)] \in \mathbb{R}^{d_a}$$

$$type(i_t) \in \{click, view, purchase, like, comment\}$$

The context feature $f_{context}$ consists of three sub-features:

- Interactive device (Device, such as PC, Mobile, Tablet)
- Interaction Platform (Platform, such as Web, iOS, Android, MiniApp, etc.)

- Geographic location (Geo embedding, such as country code)

The device type is one-hot encoded and mapped to a $d_{dev}$ dimensional embedding. The interactive platform is one-hot encoded and mapped to a $d_{plat}$ dimensional embedding. The geographic location is discretely encoded and then the $d_{geo}$ dimensional embedding is obtained by table lookup:

$$f_{context} = [e_{dev} \oplus e_{plat} \oplus e_{geo}] \in \mathbb{R}^{d_{dev}+d_{plat}+d_{geo}}$$

where $\oplus$ represent vector concatenation.

（2） Timing characteristics

Each interaction event has a timestamp $time(i_t)$, which is mapped to a time vector $t_t$ through a function $t_t = \phi(time(i_t))$. Time series features $t_t$ include:

- Absolute time $e_{abs}$
- Time interval $e_{gap}$
- Periodicity $e_{cycle}$

Absolute time $e_{abs}$: date/hour (year, month, day, hour), encoded using sine and cosine positions:

$$e_{abs}(time) = [\sin(\omega_k \cdot time), \cos(\omega_k \cdot time)], \quad k = 1, \ldots, \frac{d_{abs}}{2}$$

Time interval $e_{gap}$: $\Delta t$ will be logarithmically scaled and embedded in $\mathbb{R}^{d_{gap}}$:

$$\Delta t = \log(1 + (time(i_t) - time(i_{t-1})))$$

$$e_{gap} = Emb(\Delta t) \in \mathbb{R}^{d_{gap}}$$

Periodicity $e_{cycle}$: Weekday/weekend preferences and seasonal preferences (such as holidays) are all mapped into vectors through discrete one-hot encoding and table lookup:

$$e_{cycle} = [\sin(\tfrac{2\pi \cdot dayofweek(i_t)}{7}), \cos(\tfrac{2\pi \cdot dayofweek(i_t)}{7})]$$

The final time series feature vector is:

$$t_t = [e_{abs} \oplus e_{gap} \oplus e_{cycle}] \in \mathbb{R}^{d_{abs}+d_{gap}+d_{cycle}}$$

（3） Multimodal features

Items that are interacted with by user over time $t$ have multimodal representations $x_t^{MM}$, including:

- Text features $x_t^{text}$
- Image features $x_t^{img}$
- Video Features $x_t^{video}$
- Audio characteristics $x_t^{audio}$

Right now:

$$x_t^{MM} = \left\{ x_t^{text}, x_t^{img}, x_t^{video}, x_t^{audio} \right\}$$

Text features $x_t^{text}$ come from product details, user reviews, article descriptions, etc. The text is encoded by BERT and represented by the [CLS] vector:

$$x_t^{text} = \text{BERT}(token(i_t)) \in \mathbb{R}^{d_{text}}$$

Image features $x_t^{img}$ come from product images, social media images, and video thumbnails. The Vision Transformer (ViT) is used to extract the final layer [CLS] vector:

$$x_t^{img} = \text{ViT}(image(i_t)) \in \mathbb{R}^{d_{img}}$$

Video features $x_t^{video}$ come from product introduction videos, film clips, or short videos watched by users. We use Video Transformer (TimeSformer) to extract temporal features and perform average pooling:

$$x_t^{video} = \text{VideoTRM}(video(i_t)) \in \mathbb{R}^{d_{video}}$$

Audio features $x_t^{audio}$ are suitable for music recommendations or videos with sound. The pre-trained model wav2vec 2.0 is used to extract the latent vector:

$$x_t^{audio} = \text{Wav2Vec2.0}(audio(i_t)) \in \mathbb{R}^{d_{audio}}$$

（4） Unified feature representation

The complete representation of an interaction at a time step $t_t$ is:

$$x_t = [e_{action} \oplus f_{context} \oplus t_t \oplus x_t^{text} \oplus x_t^{img} \oplus x_t^{video} \oplus x_t^{audio}]$$

The user $u$'s complete historical interaction sequence is:

$$H_u = \{x_1, x_2, \ldots, x_t\}$$

This sequence is input to the timing function $g(\cdot)$ to obtain the dynamic state of interest:

$$z_t = g(x_1, x_2, \ldots, x_t)$$

### 5.3.3 Temporal Interest Modeling Layer

In temporal interest modeling, TIMGen relies on the self-attention mechanism of Transformer to achieve long-range dependency modeling in long sequences, avoiding the gradient vanishing problem that is prone to occur in RNN/LSTM in long temporal series modeling.

The input user history interaction sequence $H_u$ is obtained through the input encoding layer $x_1, x_2, \ldots, x_t$ and input into the Transformer's multi-head self-attention module:

$$z_t = \text{Transformer}(x_1, x_2, \ldots, x_t)$$

Here, the interest representation at each time step $z_t$ is obtained by jointly modeling the sequence context, rather than relying solely on local information. The core computational process of the Transformer is:

$$\text{Attention}(Q, K, V) = \text{softmax}\left(\frac{QK^\top}{\sqrt{d_k}}\right)V$$

The Query matrix $Q$, Key matrix $K$, and Value matrix $V$ are obtained by linearly transforming the input sequence and $d_k$ represent the dimensions of the Key vector. During time series modeling, a stepwise aggregation approach is employed, utilizing the encoding of all past interactions during calculation $z_t$. The final generation relies primarily on the most recent interest status $z_T$. This design approach aligns with real-world user interest patterns: past behaviors reflect stable, long-term interests, while the most recent interaction reflects immediate interests and preferences.

### 5.3.4 Multimodal interest fusion layer

As a supplement to the temporal interest model, the multimodal interest fusion layer calculates the multimodal interest vector in parallel with the temporal interest modeling layer. The multimodal interest fusion layer further introduces a multimodal attention mechanism to learn the contribution of each modality to the user's interest at different times. For each modal data in $m \in \{text, img, video, audio\}$, it is encoded through an independent feature extraction network (such as BERT, ViT, Video Transformer, Wav2Vec2.0). In order to ensure consistency with the user's temporal interest state $z_t$, the multimodal features are projected into a unified embedding space and then calculated as the input of the self-attention mechanism to obtain a single modal representation $h_m$ that integrates the internal contextual relationship, and then the final multimodal interest $z_t^{multi}$ is generated by weighted fusion:

$$Q_m = x_t^m W_Q^m$$

$$K_m = x_t^m W_K^m$$

$$V_m = x_t^m W_V^m$$

$$h_m = \text{Attention}(Q_m, K_m, V_m)$$

$$z_t^{multi} = \sum_m \alpha_m h_m$$

The weight $\alpha_m$ is determined by a learnable attention function, which reflects the relative contribution of different modalities to user interest at each time step $t$:

$$\alpha_m = \frac{\exp(w_m^\top h_m)}{\sum_j \exp(w_j^\top h_j)}$$

The weighting mechanism allows the model to automatically adjust the weights of modalities in specific scenarios, preventing certain modalities from being artificially assigned excessive or low importance. For example, for movie recommendations, reviews may better reflect audience interest than images, but for an e-commerce platform, images are often more persuasive than text.

**5.3.5 Personalized generation layer**

personalized generation layer is to generate personalized outputs using the interest representations derived from the previous temporal modeling and multimodal fusion. Compared to GAN-based methods, the VAE (Variational Autoencoder) has a more ordered distribution in the latent space. It can use latent variables to model complex interests and produce diverse results through sampling. Therefore, VAE is used to model the correspondence between user interests and target content in the latent space, capturing the uncertainty and diversity in user interest representations.

Further integrate temporal interests $z_t$ and multimodal interests $z_t^{multi}$ into:

$$z_t^{final} = W[z_t \oplus z_t^{multi}]$$

fused interest representation $z_t^{final}$ is mapped to the latent space, and then the latent variable $\ell$ is sampled through the reparameterization trick:

$$\mu = W_\mu z_t^{final} + b_\mu$$

$$\sigma = softplus\left(W_\sigma z_t^{final} + b_\sigma\right)$$

$$\ell = \mu + \sigma \odot \epsilon, \quad \epsilon \sim \mathcal{N}(0, I)$$

Where is $\mu$ the mean vector of the latent variable, $\sigma$ is the standard deviation vector of the latent variable, is guaranteed to be non-negative by softplus, and is learned by the parameterized network. In the decoding stage, the latent variable $\ell$ is input to the generator to generate personalized content

$$\hat{y} = G(\ell, x^*)$$

The training objective consists of two parts: the reconstruction loss, which measures the consistency between the generated content and the real content; and the KL divergence regularization term, which constrains the consistency between the latent variable distribution and the prior distribution. The objective function of the generation layer is:

$$\mathcal{L}_{VAE} = \mathbb{E}_{q_\phi(\ell|z_t^{final})}[\log p_\theta(x|\ell)] - KL(q_\phi(\ell|z_t^{final}) \| p(\ell))$$

The first term ensures that the generated results are consistent with the user's true interests, and the second term ensures the stability and generalizability of the latent variable distribution. This objective function will be added to the final loss function for optimization.

### 5.3.6 Output layer

The output layer is responsible for generating the final personalized content $\hat{y}$ based on the user's fused interest representation $z_t^{final}$ and the multimodal features $x^*$ of the candidate items, while ensuring that the generated content is consistent with the user's personalized rating and category labels. The generative model generates content of interest to the user by sampling latent variable $\ell$ and using a decoder. The output content must meet both rating accuracy constraints and category matching constraints.

generate personalized content $\hat{y}$ based on the user's interest status $z_t^{final}$ and the multimodal features of the candidate items $x^*$:

$$\hat{y} = G(z_t^{final}, x^*)$$

Compared and optimized with the user's personalized rating labels $y_t^{score}$ and category labels $y_t^{class}$. The loss function optimizes the consistency of the two types of labels simultaneously, that is, it guides the generation process by minimizing the difference between the generated content and the true label:

$$\mathcal{L}_{output} = \mathcal{L}_{VAE} + \lambda_1 \mathcal{L}_{score} + \lambda_2 \mathcal{L}_{class}$$

$$\mathcal{L}_{score} = \| \hat{y}^{score} - y^{score} \|^2$$

$$\mathcal{L}_{class} = -\sum_{k=1}^{K} y_k^{class} \log \hat{y}_k^{class}$$

where $\mathcal{L}_{score}$ is used to measure the difference between generated content and personalized scoring labels, $\mathcal{L}_{class}$ is used to measure the match between generated content and category labels.

# 6 Discussions and Future Work

TIMGen framework combines temporal interest modeling, multimodal fusion, and generative methods to capture the dynamic nature and modal diversity of user interests, while also leveraging dual labeling to constrain personalization and relevance. At the application level, the framework offers strong scalability, enabling applications such as education, healthcare, and social platforms to generate personalized learning resources, health recommendations, or social content.

it can capture long-term interests and trends through explicit time embedding and a Transformer-based long-term sequence mechanism; (2) it uses the attention mechanism to assign weights to text, images, videos, and audio, allowing for flexible adjustment and diverse output; and (3) it uses a combination of rating and category labels to simultaneously improve personalization and relevance.

The shortcomings of TIMGen are: (1) Although the collected data introduces different modal information such as video and audio, the data annotation is unbalanced and cannot fully capture all information and modalities. (2) Generation relies on the features of external candidate items $x^*$ and is not completely new content. (3) Although the long-term time series modeling based on Transformer can better reflect historical dependencies, recent interaction signals are easily diluted by historical information. In the face of some scenarios with high requirements for instantaneous interest fluctuations (such as the immediacy of real-time advertising), it still has certain limitations. (4) Although the multimodal fusion mechanism achieves a certain degree of dynamic adaptability, it is difficult to capture complex and high-order modal interactions. (5) The generation model based on VAE is inferior to GANs and Diffusion in terms of visual quality. If the generation effect is high, such as wanting to generate a large image product, VAE is not the best choice.

Future improvement directions include: (1) further constructing larger-scale and more balanced multimodal long-time series datasets ; (2) introducing hierarchical interest modeling, while maintaining the ability to learn changes in long-time series , introducing specific mechanisms to enhance the response to immediate interests; (3) using cross-attention or graph neural networks to enhance the interaction between and within modalities ; (4) combining VAE with GANs/Diffusion to improve generation quality, and using large-scale pre-training generation to improve generalization ability ; (5) adding more labels and categories, and injecting information such as sequence preferences and explicit feedback to enrich the content of the supervision signal and improve its granularity .